\documentstyle[preprint,tighten,prc,aps,epsf]{revtex}
\begin{document}
\def\simge{\hspace*{0.2em}\raisebox{0.5ex}{$>$}
     \hspace{-0.8em}\raisebox{-0.3em}{$\sim$}\hspace*{0.2em}}
\def\simle{\hspace*{0.2em}\raisebox{0.5ex}{$<$}
     \hspace{-0.8em}\raisebox{-0.3em}{$\sim$}\hspace*{0.2em}}
\def\bra#1{{\langle#1\vert}}
\def\ket#1{{\vert#1\rangle}}
\def\coeff#1#2{{\scriptstyle{#1\over #2}}}
\def\undertext#1{{$\underline{\hbox{#1}}$}}
\def\hcal#1{{\hbox{\cal #1}}}
\def\sst#1{{\scriptscriptstyle #1}}
\def\eexp#1{{\hbox{e}^{#1}}}
\def\rbra#1{{\langle #1 \vert\!\vert}}
\def\rket#1{{\vert\!\vert #1\rangle}}
\def\lsim{{ <\atop\sim}}
\def\gsim{{ >\atop\sim}}
\def\nubar{{\bar\nu}}
\def\psibar{{\bar\psi}}
\def\Gmu{{G_\mu}}
\def\alr{{A_\sst{LR}}}
\def\wpv{{W^\sst{PV}}}
\def\evec{{\vec e}}
\def\notq{{\not\! q}}
\def\notl{{\not\! \ell}}
\def\notk{{\not\! k}}
\def\notp{{\not\! p}}
\def\notpp{{\not\! p'}}
\def\notder{{\not\! \partial}}
\def\notcder{{\not\!\! D}}
\def\notA{{\not\!\! A}}
\def\notv{{\not\!\! v}}
\def\Jem{{J_\mu^{em}}}
\def\Jana{{J_{\mu 5}^{anapole}}}
\def\nue{{\nu_e}}
\def\mn{{M_\sst{N}}}
\def\mns{{M^2_\sst{N}}}
\def\me{{m_e}}
\def\mes{{m^2_e}}
\def\mq{{m_q}}
\def\mqs{{m_q^2}}
\def\mz{{M_\sst{Z}}}
\def\mzs{{M^2_\sst{Z}}}
\def\ubar{{\bar u}}
\def\dbar{{\bar d}}
\def\sbar{{\bar s}}
\def\qbar{{\bar q}}
\def\sstw{{\sin^2\theta_\sst{W}}}
\def\gv{{g_\sst{V}}}
\def\ga{{g_\sst{A}}}
\def\pv{{\vec p}}
\def\pvs{{{\vec p}^{\>2}}}
\def\ppv{{{\vec p}^{\>\prime}}}
\def\ppvs{{{\vec p}^{\>\prime\>2}}}
\def\qv{{\vec q}}
\def\qvs{{{\vec q}^{\>2}}}
\def\xv{{\vec x}}
\def\xpv{{{\vec x}^{\>\prime}}}
\def\yv{{\vec y}}
\def\tauv{{\vec\tau}}
\def\sigv{{\vec\sigma}}
\def\sst#1{{\scriptscriptstyle #1}}
\def\gpnn{{g_{\sst{NN}\pi}}}
\def\grnn{{g_{\sst{NN}\rho}}}
\def\gnnm{{g_\sst{NNM}}}
\def\hnnm{{h_\sst{NNM}}}
\def\xivz{{\xi_\sst{V}^{(0)}}}
\def\xivt{{\xi_\sst{V}^{(3)}}}
\def\xive{{\xi_\sst{V}^{(8)}}}
\def\xiaz{{\xi_\sst{A}^{(0)}}}
\def\xiat{{\xi_\sst{A}^{(3)}}}
\def\xiae{{\xi_\sst{A}^{(8)}}}
\def\xivtez{{\xi_\sst{V}^{T=0}}}
\def\xivteo{{\xi_\sst{V}^{T=1}}}
\def\xiatez{{\xi_\sst{A}^{T=0}}}
\def\xiateo{{\xi_\sst{A}^{T=1}}}
\def\xiva{{\xi_\sst{V,A}}}

\def\rvz{{R_\sst{V}^{(0)}}}
\def\rvt{{R_\sst{V}^{(3)}}}
\def\rve{{R_\sst{V}^{(8)}}}
\def\raz{{R_\sst{A}^{(0)}}}
\def\rat{{R_\sst{A}^{(3)}}}
\def\rae{{R_\sst{A}^{(8)}}}
\def\rvtez{{R_\sst{V}^{T=0}}}
\def\rvteo{{R_\sst{V}^{T=1}}}
\def\ratez{{R_\sst{A}^{T=0}}}
\def\rateo{{R_\sst{A}^{T=1}}}
\def\mro{{m_\rho}}
\def\mks{{m_\sst{K}^2}}
\def\mpi{{m_\pi}}
\def\mpis{{m_\pi^2}}
\def\mom{{m_\omega}}
\def\mphi{{m_\phi}}
\def\Qhat{{\hat Q}}
\def\FOS{{F_1^{(s)}}}
\def\FTS{{F_2^{(s)}}}
\def\GAS{{G_\sst{A}^{(s)}}}
\def\GES{{G_\sst{E}^{(s)}}}
\def\GMS{{G_\sst{M}^{(s)}}}
\def\GATEZ{{G_\sst{A}^{\sst{T}=0}}}
\def\GATEO{{G_\sst{A}^{\sst{T}=1}}}
\def\mdax{{M_\sst{A}}}
\def\mustr{{\mu_s}}
\def\rsstr{{r^2_s}}
\def\rhos{{\rho_s}}
\def\GEG{{G_\sst{E}^\gamma}}
\def\GEZ{{G_\sst{E}^\sst{Z}}}
\def\GMG{{G_\sst{M}^\gamma}}
\def\GMZ{{G_\sst{M}^\sst{Z}}}
\def\GEn{{G_\sst{E}^n}}
\def\GEp{{G_\sst{E}^p}}
\def\GMn{{G_\sst{M}^n}}
\def\GMp{{G_\sst{M}^p}}
\def\GAp{{G_\sst{A}^p}}
\def\GAn{{G_\sst{A}^n}}
\def\GA{{G_\sst{A}}}
\def\GETEZ{{G_\sst{E}^{\sst{T}=0}}}
\def\GETEO{{G_\sst{E}^{\sst{T}=1}}}
\def\GMTEZ{{G_\sst{M}^{\sst{T}=0}}}
\def\GMTEO{{G_\sst{M}^{\sst{T}=1}}}
\def\lamd{{\lambda_\sst{D}^\sst{V}}}
\def\lamn{{\lambda_n}}
\def\lams{{\lambda_\sst{E}^{(s)}}}
\def\bvz{{\beta_\sst{V}^0}}
\def\bvo{{\beta_\sst{V}^1}}
\def\Gdip{{G_\sst{D}^\sst{V}}}
\def\GdipA{{G_\sst{D}^\sst{A}}}
\def\fks{{F_\sst{K}^{(s)}}}
\def\FIS{{F_i^{(s)}}}
\def\fpi{{f_\pi}}
\def\fk{{f_\sst{K}}}
\def\RAp{{R_\sst{A}^p}}
\def\RAn{{R_\sst{A}^n}}
\def\RVp{{R_\sst{V}^p}}
\def\RVn{{R_\sst{V}^n}}
\def\rva{{R_\sst{V,A}}}
\def\xbb{{x_B}}
\def\mlq{{M_\sst{LQ}}}
\def\mlqs{{M_\sst{LQ}^2}}
\def\lscal{{\lambda_\sst{S}}}
\def\lvect{{\lambda_\sst{V}}}

\def\PR#1{{{\em   Phys. Rev.} {\bf #1} }}
\def\PRC#1{{{\em   Phys. Rev.} {\bf C#1} }}
\def\PRD#1{{{\em   Phys. Rev.} {\bf D#1} }}
\def\PRL#1{{{\em   Phys. Rev. Lett.} {\bf #1} }}
\def\NPA#1{{{\em   Nucl. Phys.} {\bf A#1} }}
\def\NPB#1{{{\em   Nucl. Phys.} {\bf B#1} }}
\def\AoP#1{{{\em   Ann. of Phys.} {\bf #1} }}
\def\PRp#1{{{\em   Phys. Reports} {\bf #1} }}
\def\PLB#1{{{\em   Phys. Lett.} {\bf B#1} }}
\def\ZPA#1{{{\em   Z. f\"ur Phys.} {\bf A#1} }}
\def\ZPC#1{{{\em   Z. f\"ur Phys.} {\bf C#1} }}
\def\etal{{{\em   et al.}}}

\def\delalr{{{delta\alr\over\alr}}}
\def\pbar{{\bar{p}}}
\def\lamchi{{\Lambda_\chi}}

\def\qw0{{Q_\sst{W}^0}}
\def\qwp{{Q_\sst{W}^P}}
\def\qwn{{Q_\sst{W}^N}}
\def\qwe{{Q_\sst{W}^e}}
\def\qem{{Q_\sst{EM}}}

\def\gae{{g_\sst{A}^e}}
\def\gve{{g_\sst{V}^e}}
\def\gvf{{g_\sst{V}^f}}
\def\gaf{{g_\sst{A}^f}}
\def\gvu{{g_\sst{V}^u}}
\def\gau{{g_\sst{A}^u}}
\def\gvd{{g_\sst{V}^d}}
\def\gad{{g_\sst{A}^d}}

\def\gvftil{{\tilde g_\sst{V}^f}}
\def\gaftil{{\tilde g_\sst{A}^f}}
\def\gvetil{{\tilde g_\sst{V}^e}}
\def\gaetil{{\tilde g_\sst{A}^e}}
\def\gvqtil{{\tilde g_\sst{V}^e}}
\def\gaqtil{{\tilde g_\sst{A}^e}}
\def\gvutil{{\tilde g_\sst{V}^e}}
\def\gautil{{\tilde g_\sst{A}^e}}
\def\gvdtil{{\tilde g_\sst{V}^e}}
\def\gadtil{{\tilde g_\sst{A}^e}}

\def\hvf{{h_\sst{V}^f}}
\def\hvu{{h_\sst{V}^u}}
\def\hvd{{h_\sst{V}^d}}
\def\hve{{h_\sst{V}^e}}
\def\hvq{{h_\sst{V}^q}}

\def\delp{{\delta_P}}
\def\delzp{{\delta_{00}}}
\def\deld{{\delta_\Delta}}
\def\dele{{\delta_e}}

\def\apv{{A_\sst{PV}}}
\def\apvnsid{{A_\sst{PV}^\sst{NSID}}}
\def\apvnsd{{A_\sst{PV}^\sst{NSD}}}
\def\qpv{{Q_\sst{W}}}

\def\lamchi{{\Lambda_\chi}}
\def\lamchis{{\Lambda_\chi^2}}
\def\lamchic{{\Lambda_\chi^3}}
\def\FOa{{F_1^{(a)}}}
\def\FTa{{F_2^{(a)}}}
\def\GAS{{G_\sst{A}^{(s)}}}
\def\GEa{{G_\sst{E}^{(a)}}}
\def\GMa{{G_\sst{M}^{(a)}}}
\def\c{{{\cal C}}}
\def\muC{{\tilde{\mu}_{\mbox{{\tiny C}}}}}
\def\muT{{\tilde{\mu}_{\mbox{{\tiny T}}}}}
\def\mus{{\mu_s}}
\def\barq{{\bar q}}
\def\bars{{\bar s}}
\def\bo{{b_0}}
\def\bp{{b_+}}
\def\bm{{b_-}}
\def\bix{{b_9}}
\def\bx{{b_{10}}}
\def\bxi{{b_{11}}}
\def\co{{c_0}}
\def\cp{{c_+}}
\def\cm{{c_-}}
\def\musc{{\tilde{\mu}^s_{\mbox{{\tiny C}}}}}
\title{Octet Baryon Charge Radii, Chiral Symmetry and Decuplet Intermediate
States}

\author{S.J. Puglia$^a$ \\
M.J. Ramsey-Musolf$^{a,b}$\\
Shi-Lin Zhu$^a$}
\address{
$^a$ Department of Physics, University of Connecticut,
Storrs, CT 06269 USA \\
$^b$Theory Group, Thomas Jefferson National Accelerator Facility, Newport News,
VA 23606 }

\maketitle
\begin{abstract}
We compute the octet baryon charge radii to ${\cal O}(1/\lamchis\mn)$ in
heavy baryon chiral
perturbation theory (HB$\chi$PT).
We examine the effect of including the decuplet of spin-$3\over2$ baryons
explicitly.
We find that it does not
improve the level of agreement between the HB$\chi$PT and experimental
values for the $\Sigma^-$ charge radius.
\end{abstract}
\bigskip

The study of electromagnetic (EM) form factors is an important tool used
in developing a theoretical understanding of the internal structure of
hadrons.
In particular, the application of chiral perturbation theory ($\chi$PT) to
this study
allows one to disentangle the long- and short-range QCD dynamics governing
the behavior of
form factors at low momentum transfers. The long-range effects appear in
the guise of
non-analytic loop contributions while the short-range dynamics are subsumed
into
phenomenologically determined low-energy constants (LEC's).

An important issue in the viability 
of this program is the convergence of the chiral expansion. 
An important consideration in obtaining proper covergence is 
the treatment of the large baryon mass. In many cases, one 
obtains adequate convergence by employing a non-relativistic version 
of $\chi$PT, involving a second expansion in inverse powers of the 
baryon mass. In addition, 
it has been 
found that for some 
observables, explicit inclusion of the decuplet of spin-$3\over2$ baryons 
is necessary in order 
to achieve the expected convergence behavior. For octet of axial currents 
\cite{JM2}, for example, 
the contribution of the decuplet is comparable to that of the octet and 
opposite in sign, 
leading to sizeable cancellations. In the case of the octet baryon 
magnetic moments, this convergence has been a subject of controversy 
\cite{eg2}\cite{MS2}. 
Recently, however, it was shown that inclusion of decuplet intermediate 
states -- together 
with the full set of $1/\mn$ corrections -- produces the expected behavior 
of the expansion \cite{SPMRM}. 
 
In contrast, the behavior of the corresponding expansion for the baryon 
octet charge radii has not been 
extensively studied. Existing calculations have worked only to leading 
order in $1/\lamchi$, where 
$\lamchi=4\pi F_\pi\sim 1$ GeV is the scale of chiral symmetry breaking 
\cite{MS1}\cite{TC}. The recent 
SELEX measurement of the $\Sigma^-$ charge radius \cite{sel} makes further 
investigation of the 
chiral convergence for the octet radii timely. In this paper, we address 
this issue by computing the 
radii through sub-leading order. As in the case of the magnetic moments, we 
consider the impact of 
explicit inclusion of decuplet intermediate states as well as $1/\mn$ 
corrections. We find that at 
order $1/\mbox{Heavy}^3$, where \lq\lq Heavy" denotes either $\lamchi$ or 
$\mn$, the chiral 
expansion for the charge radii does not appear to be under control. 
Moreover, attaining further progress 
is hampered by the lack of available data. Although a computation of the 
next order non-analytic contributions 
feasible, the new LEC's arising at this order could not be determined from 
experiment. Thus, when baryons are treated in a non-relativistic framework, a 
chiral analysis of baryon charge radii is viable only for the nucleon 
isovector radius in the context of SU(2)\footnote{For an analsis of nucleon 
EM form factors in relativistic baryon $\chi$PT, see Ref. \cite{MS3}.}. 
 
$\chi$PT is an effective field theory of the strong interactions which 
exploits the approximate 
chiral symmetry of the QCD Lagrangian. The spontaneous breakdown of this 
symmetry gives rise to 
eight (nearly) massless Goldstone bosons. These are identified with the 
octet of pseudoscalar mesons. 
In the meson sector a well-defined low-energy expansion in terms of the 
meson 4-momentum can be 
constructed to describe the dynamics. Here low-energy is defined relative 
to the scale of chiral symmetry 
breaking $\lamchi= 4\pi F_\pi\approx 1$ GeV, where 
$F_\pi$ is the pion decay constant. The baryons can be included in a 
consistent manner as 
shown in Ref.\cite{JM2} The non-relativistic version of 
$\chi$PT including the baryons -- 
heavy baryon chiral perturbation theory (HB$\chi$PT) -- has many 
simplifying features. A 
review of these features can be found in Ref. \cite{JM1}. HB$\chi$PT has 
proven useful in 
the calculation of low-energy observables such as baryon masses, Compton 
scattering amplitudes, 
nucleon polarizabilities, sigma terms, axial cuurent, and hyperon decays 
\cite{eg}. The important 
point for the present study is that HB$\chi$PT introduces a second 
expansion scale $\mn$ commensurate 
with $\lamchi$. Thus, one should include all contributions to a given order 
in $1/{\hbox{Heavy}}$, 
where \lq\lq Heavy" denotes either $\lamchi$ or $\mn$. 
In Ref. \cite{MS2} three-flavor HB$\chi$PT was used to calculate the EM 
form fatcors for the octet. 
The analysis there was carried out to ${\cal{O}}(1/\mbox{Heavy}^2)$ and did 
not include the decuplet as 
 an explicit degree of freedom. Our approach differs from theirs in two 
respects: 1.) Inclusion of 
the decuplet, 2.) inclusion of $1/(\mn)$ corrections. 
 
The formalism of HB$\chi$PT is by now well known and we refer the reader to 
Ref. \cite{SPMRM} for 
our conventions and notation. As in Ref.\cite{MS2} we define the charge 
radius by 
\begin{equation} 
\langle r^2_E \rangle=\left. {1\over N} \frac{dG_E(q^2)}{dq^2} \right|_{q^2=0} 
\end{equation} 
where $G_E$ is the Sachs electric form factor and $N$ is a normalization 
constant. 
The  tree-level contributions to the charge 
radii are ${\cal{O}}(1/\lamchis)$ 
in the chiral expansion and are generated by the Lagrangian \cite{MRM/ITO} 
 
\begin{eqnarray} 
{\cal L}_{2} &=& -\frac{e}{\lamchis}\left\{c_{+}\hbox{ 
Tr}\left({\bar{B}}_v\{Q,B_v\}\right) 
 +c_{-}\hbox{ Tr}\left({\bar{B}}_v [Q,B_v]\right)\right\}v_\mu \partial_\nu 
F^{\mu\nu}. 
 \label{eq:L2} 
\end{eqnarray} 
There is an additional tree-level contribution at $1/(\mbox{Heavy})^2$ that 
arises from the $1/\mn$ 
correction of the kinetic term of the lowest order  Lagrangian and is given by 
\begin{equation} 
{\cal L}_{\frac{1}{\mn^2}}=-\frac{1}{4\mn^2}\hbox{ Tr}\left({\bar B}_v[ 
\not\!\!D^{\bot},[ v\cdot D, [\not\!\!D^{\bot},B_v ]]] 
\right\}, 
\label{eq:kin} 
\end{equation} 
where $D^{\bot}_\mu = D_\mu-v_\mu v\cdot D$. 
 
The relevant one-loop graphs which arise at order ${\cal O}(1/\lamchis)$ 
are shown 
in Fig. 1 (b-f), where the vertices arise from the Lagrangian:\cite{Hem} 
 
\begin{eqnarray} 
{\cal L}_0&=&i\hbox{ Tr}\left({\bar B}_vv\cdot D\ B_v\right)+ 2D 
\hbox{ Tr}\left({\bar B}_v S_v^\mu\{A_\mu, B_v\}\right) 
+2F\hbox{ Tr}\left({\bar B}_v S_v^\mu[A_\mu, B_v]\right)\nonumber \\ 
& &  -i{\bar T}^\mu_v \left(v\cdot 
{\cal D}\right)T_{v\mu} 
+\delta {\bar T}^\mu_v T_{v\mu}+\c\left({\bar T}^\mu_v A_\mu B_v+{\bar 
B}_v A_\mu T_{v\mu}\right)+ 2{\cal H}{\bar T}^\mu_v S_v^\nu A_\nu 
T_{v\mu}\nonumber \\ 
& &+{\fpi^2\over4}\hbox{Tr}\left((D^\mu\Sigma)^{\dag}D_\mu\Sigma\right). 
\label{eq:lheav} 
\end{eqnarray} 
In the following we take $D=.75$, $F=.50$ and ${\cal C}=-1.5$ 
 
The loop corrections to the charge radius at $1/(\mbox{Heavy})^2$ are 
represented by the graphs (c-g) 
in Fig. 1. The contributions to the charge radii from (d),(e) and (g) 
vanish indentically since these 
graphs have no  $q^2$ dependence at this order in the chiral expansion. 
With the inclusion of the counterterm and the $1/\mn^2$ correction, the 
charge radii have the form: 
\begin{equation} 
\langle r^2_B\rangle= {6N\over\lamchis} \left\{\alpha_B-\sum_{X=\pi,K}\left( 
 \beta^{(X)}_B \ln\frac{m_X^2}{\mu^2}+\beta^{\prime(X)}_B 
 F(m_X,\delta,\mu)\right)\right\}+{3\tilde{\alpha}_B\over2\lamchi\mn}-{3Q_B\over 
4\mn^2}, 
 \label{eq:r1} 
 \end{equation} 
Here $N$ is a normalization factor which is equal to the charge of the 
baryon if it is charged or 1 
otherwise and $Q_B$ is the 
 charge of the baryon . The coefficients $\alpha_B$ are linear 
combinations of the counterterm 
couplings 
$c_\pm$ from Eq.(\ref{eq:L2}), while the $\beta^{(X)}_B$ and 
$\beta^{\prime(X)}_B$ are products of the 
meson charge and Clebsch-Gordon coefficients arising from the one-loop 
graphs in Fig. 1. The second to last term is the so called \lq\lq Foldy"-term 
\cite{MS3} and arises from the definition of the $G_E$. This term depends 
on two magnetic couplings which we call $b_\pm$ and are contained in the 
$\tilde{\alpha}_B$'s \cite{MS2,MS3}. The last term in Eq.(\ref{eq:r1}) is 
the relativistic correction from 
Eq.(\ref{eq:kin}). 
 
We note  that the coefficients, $\beta^{\prime(X)}_B$, for the graph 
containing the decuplet have not been computed before. All of these 
coefficients are listed in the Appendix. 
The analytic structure for this graph is given by: 
\begin{equation} 
F\left(m,\delta,\mu\right)=\ln\frac{m^2}{\mu^2}+2\delta\left\{ 
\begin{array}{ll} 
\frac{1}{\sqrt{m^2-\delta^2}}\left(\frac{\pi}{2}- 
\arctan\left[\frac{\delta}{\sqrt{m^2-\delta^2}}\right]\right)& 
 m\geq \delta\nonumber \\ 
  & \nonumber \\ 
 -\frac{1}{\sqrt{\delta^2-m^2}}\ln\left[\frac{\delta+ 
 \sqrt{\delta^2-m^2}}{m}\right]& 
 m<\delta\ , 
 \end{array} 
 \right. 
\end{equation} 
where $\delta$ is the octet-decuplet mass splitting and $\mu$ is the 
renormalization scale. 
 
The contributions at ${\cal O}(1/\mbox{Heavy}^3)$ are all proportional to 
$1/\lamchis\mn$. There are no 
terms proprtional to $1/\lamchic$ with the correct spacetime structure and 
terms proportional 
to $1/\lamchi\mn^2$ or 
$1/\mn^3$ will vanish at $q^2=0$. The terms proportional to $1/\lamchis\mn$ 
are of two types. 
First the graphs in Fig. 1 can be expanded as functions of $q^2$ and it is 
found that they 
contain contributions proportional to $q^2/2\mn$. The graphs (c), (d), (f) 
and (g) are of this 
type and will contribute to the charge radius at this order. There are also 
graphs that contain an 
explicit factor of $1/\mn$ which arises for the heavy baryon expansion. 
These are the graphs shown in 
Fig. 2, which contain vertices generated from the Lagragian \cite{MS2} 
\begin{eqnarray} 
{\cal L}_{\mn} & = & 
\frac{1}{2\mn}\hbox{ Tr}\left(\bar{B}\left[v\cdot D,\big[v\cdot D,B\big]\right] 
-\bar{B}\left[D^\mu,\left[D_\mu,B\right]\right]\right) 
\nonumber \\[.5em] 
& & 
-\frac{i\,D}{2\mn} 
\hbox{ Tr}\left( 
\bar{B}S_\mu\left[D^\mu,\left\{v\cdot A,B\right\}\right] 
+\bar{B}S_\mu\left\{v\cdot A,\left[D^\mu,B\right]\right\} 
\right)\nonumber \\[.5em] 
& & 
-\frac{i\,F}{2\mn} 
\hbox{ Tr}\left( 
\bar{B}S_\mu\left[D^\mu,\left[v\cdot A,B\right]\right] 
+\langle\bar{B}S_\mu\left[v\cdot A,\left[D^\mu,B\right]\right] 
\right) \nonumber\\[.5em] 
& & 
+\frac{1}{2\mn}{\bar T}^\mu_v\left({\cal D}^\alpha {\cal D}_\alpha-v\cdot {\cal 
D}\ v\cdot {\cal 
D}\right)T_{v\mu} \,\,\, . 
\end{eqnarray} 
 
Although there are thirteen graphs which appear at ${\cal 
O}(1/\mbox{Heavy}^3)$, the contributions 
from several vanish. 
The amplitude of Fig 1. (e) again has no $q^2$ dependence. 
The $1/\mn$ contribution from Fig 1. (d) combines with those of Fig 2. (c) 
and (d) to sum to zero. A similar cancellation takes place between the 
$1/\mn$ correction to Fig. 1 
(g) and Fig. 2 (e) and (f). The graphs in Fig. 2 (g) and (h) vanish at this 
order. 
Thus, only those in Fig. 1 (c) and (f) and Fig. 2 (a) and (b) actually 
end up contributing to the charge radius. 
Taking into account these contributions the charge radius now has the form: 
 
\begin{eqnarray} 
\langle r^2_B\rangle&=& {6N\over\lamchis} \left\{\alpha_B-\sum_{X=\pi,K}\left( 
 \beta^{(X)}_B \left(\ln\frac{m_X^2}{\mu^2}+{23\pi 
m_X\over24\mn}\right)+\beta^{\prime(X)}_B 
 \left(F(m_X,\delta,\mu)-{2\over\mn}G(m_X,\delta,\mu)\right)\right)\right\}\nonumber\\ 
 & & +{3\tilde{\alpha}_B\over2\lamchi\mn}-{3Q_B\over4\mn^2}, 
 \label{eq:r2} 
 \end{eqnarray} 
where 
\begin{equation} 
G\left(m,\delta,\mu\right)=\delta\ln\frac{m^2}{\mu^2}+(2\delta^2-m^2)\left\{ 
\begin{array}{ll} 
\frac{1}{\sqrt{m^2-\delta^2}}\left(\frac{\pi}{2}- 
\arctan\left[\frac{\delta}{\sqrt{m^2-\delta^2}}\right]\right)& 
 m\geq \delta\nonumber \\ 
  & \nonumber \\ 
 \frac{1}{\sqrt{\delta^2-m^2}}\ln\left[\frac{\delta+ 
 \sqrt{\delta^2-m^2}}{m}\right]& 
 m<\delta\ . 
 \end{array} 
 \right. 
\end{equation} 
 
We now turn to the determination of the LEC's $c_\pm$. 
First, we need values for the magnetic couplings $b_{\pm}$ which appear in 
the Foldy term. 
These are obtained from a fit of the octet anomalous magnetic moments 
\cite{SPMRM} and are given in Table I for the cases considered below. With 
these, we use the measured values of the nucleon charge radii to determine 
$c_\pm$. We consider two cases: 1.) octet intermediate 
states only (O),  2.) both octet and decuplet intermediate states included 
(O+D). The values for the 
counterterm couplings and charge radii in each case are shown in Tables II 
and III respectively.

We see that the effect of the decuplet is large and, except for $\Xi^0$ and 
$\Lambda$, tends to increase 
the value of $\langle r^2_B\rangle$. In the case of the $\Sigma^-$ at 
${\cal O}(1/\lamchis)$ both 
the O and O+D calculations produce a value for the charge radius that is 
much larger than the SELEX 
measurement of $\langle r^2_{\Sigma^-}\rangle=0.60\pm 0.08$. The $1/\mn$ 
corrections tend to reduce this result, 
with the ${\cal O}(1/\lamchis\mn)$ octet only value being very close to the 
measured value. Here again, however, 
the addition of the decuplet increases the predicted value away from the 
measured one. We also note the 
fluctuations in sign for the calculated values for $\Xi^{0}$, $\Lambda$ and 
the $\Sigma^0\Lambda$ transition 
moment. 
 
The effect of including both the decuplet and the $1/\mn$ corrections on 
the LEC's is also dramatic. While in all cases $c_\pm$ are of natural size, 
they change 
considerably, both in sign and in magnitude, between orders and with the 
addition of the decuplet. No clear 
pattern emerges and the values do not appear to be converging. 
 
We contrast the above behavior to the magnetic moment case as analyzed in 
Ref.\cite{SPMRM}. 
There, inclusion of both the decuplet and the $1/(\mbox{Heavy}^3)$ terms 
played a essential role obtaining 
the proper convergence behavior of the chiral expansion. In that case, it 
was possible to work two 
orders beyond tree-level since there is ample data to fit higher order 
counterterms and no two-loop 
contributions. For the charge radii, however, the lowest order loop and 
tree-level contributions arise 
at the same order. Moreover, the exists considerably less data for the 
charge radii than for the magnetic 
moments. Consequently, it is impossible to analyze the radii beyond 
sub-leading order in HB$\chi$PT without 
introducing strong model assumptions. Indeed, at next-to-next-to-leading 
order, one must compute the 
full set of two-loop contributions and include a plethora of new LEC's. The 
lack of sufficient data would 
require the use of a model to determine the latter. 
 
In short, we conclude that the heavy baryon chiral expansion of the baryon 
octet charge radii is not 
yet under control. One must include the octet radii in a growing list of 
observables for which the 
convergence of the non-relativistic chiral expansion appears problematic at 
best (see, {\em e.g.}, Refs. 
\cite{Mei97,Hem00,Kum00}.) In this regard, a reanalysis of the octet 
magnetic moments and charge radii may prove 
more successful when a relativistic framework is adopted \cite{MS3}. It 
remains to be seen whether such a 
framework sufficiently reduces the number of LEC's incurred in the heavy 
baryon expansion and circumvents the 
need to work beyond sub-leading order in analyzing octet EM moments.

\begin{table}
\begin{tabular}{|c|c|c|c|c|}\hline
 & \multicolumn{2}{c |}{${\cal O}(1/\lamchis)$}&\multicolumn{2}{c |}{${\cal
O}(1/\lamchis\mn)$} \\ \hline
 $CT$ &  O    &O+D &  O    &O+D\\ \hline
  & & & &     \\
  $b_{+}$&  $2.999$ & $3.61$& $1.975$& $3.104$ \\
  $b_{-}$&  $1.571$ & $1.571$&$1.042$ &$1.393$  \\
   & &  & &\\ \hline
\end{tabular} 

\caption{Magnetic couplings appearing in the Foldy-term of the charge radius.}

\begin{tabular}{|c|c|c|c|c|}\hline
 & \multicolumn{2}{c |}{${\cal O}(1/\lamchis)$}&\multicolumn{2}{c |}{${\cal
O}(1/\lamchis\mn)$} \\ \hline
 $CT$ &  O    &O+D &  O    &O+D\\ \hline
  & & & &     \\
  $c_{+}$&  $-4.089$ & $-4.620$& $0.202$& $0.286$ \\
  $c_{-}$&  $1.410$ & $-0.211$&$2.888$ &$1.051$  \\
   & &  & &\\ \hline
\end{tabular} 

\caption{Couplings for leading order Charge Radius Counterterm.}

\begin{tabular}{|c|c|c|c|c|}\hline
  &\multicolumn{4}{c |}{$\langle r^2_B\rangle$ (fm$^2$)} \\  \hline
   & \multicolumn{2}{c |}{${\cal O}(1/\lamchis)$}&\multicolumn{2}{c
|}{${\cal O}(1/\lamchis\mn)$} \\ \hline
  &  O    &O+D&  O    &O+D\\ \hline
  & & & &    \\
  $p$& $0.735$ & $0.735$& $0.735$& $0.735$ \\
  $n$& $-0.113$ & $-0.113$&$-0.113$ &$-0.113$  \\
  $\Xi^{0}$& $ 0.251$ & $-0.001$&$0.112$ &$-0.122$  \\
  $\Xi^{-}$& $0.599$ & $0.851$ & $0.763$&$0.997$ \\
  $\Sigma^{+}$& $0.647$ & $1.522$&$0.781$ & $1.366$ \\
  $\Sigma^{-}$& $0.851$ & $0.977$&$0.681$ &$0.798$  \\
  $\Lambda$ & $0.102$ & $ -0.150$&$-0.050$ &$-0.284$ \\
  $\Sigma^0\Lambda$&$-0.021$& $-0.021$&$0.074$&$0.074$  \\
   & & & & \\ \hline
\end{tabular} 

\caption{Calculated values Charge Radii from fit of Counterterms to Nucleon experimental values.}
\end{table}

\newpage
\begin{center}
{\bf \Large APPENDIX}
\end{center}

Here we tabulate the coeffcients appearing the expressions for the
charge radii.\\

\begin{center}
\begin{tabular}{|l|l|}\hline
&\multicolumn{1}{|c|}{$\alpha_B$}\\[1mm] \hline
$p$&${1\over3}\cp+\cm$\\[2mm]
$n$&$-{2\over3}\cp$\\[2mm]
$\Xi^-$&$-{1\over3}\cp-\cm$\\[2mm]
$\Xi^0$&${-2\over3}\cp$\\[2mm]
$\Sigma^+$&${1\over3}\cp+\cm$\\[2mm]
$\Sigma^-$&${1\over3}\cp-\cm$\\[2mm]
$\Lambda$&$-{1\over3}\cp$\\[2mm]
$\Sigma^0\Lambda$&${1\over\sqrt{3}}\cp$\\[2mm]
\hline
\end{tabular}
\end{center}
\vspace*{3cm}

\begin{center}
\begin{tabular}{|l|l|}\hline
&\multicolumn{1}{|c|}{$\tilde{\alpha}_B$}\\[1mm] \hline
$p$&${1\over3}\bp+\bm$\\[2mm]
$n$&$-{2\over3}\bp$\\[2mm]
$\Xi^-$&$-{1\over3}\bp-\bm$\\[2mm]
$\Xi^0$&${-2\over3}\bp$\\[2mm]
$\Sigma^+$&${1\over3}\bp+\bm$\\[2mm]
$\Sigma^-$&${1\over3}\bp-\bm$\\[2mm]
$\Lambda$&$-{1\over3}\bp$\\[2mm]
$\Sigma^0\Lambda$&${1\over\sqrt{3}}\bp$\\[2mm]
\hline
\end{tabular}
\end{center}
\vspace*{3cm}

\begin{center}
\begin{tabular}{|l|c|c|}\hline
&\multicolumn{2}{|c|}{$\beta^{(X)}_B$}\\[1mm]  \hline
&$\pi$&$K$\\[1mm] \hline
$p$&${1\over12}+{5\over12}(D+F)^2$ &$-{1\over6}-{5\over6}({D^2\over3}+F^2)$ \\[2mm]
$n$&$-{1\over12}-{5\over6}(D+F)^2$ &${1\over12}+{5\over12}(D-F)^2$ \\[2mm]
$\Xi^-$&$-{1\over12}-{5\over12}(D-F)^2$ &$-{1\over6}-{1\over6}({2\over3}D^2+F^2)$ \\[2mm]
$\Xi^0$& ${1\over12}+{5\over12}(D-F)^2$& $-{1\over12}-{5\over12}(D+F)^2$\\[2mm]
$\Sigma^+$& ${1\over6}+{5\over6}({2\over3}D^2+F^2)$&${1\over12}+{5\over12}(D+F)^2$ \\[2mm]
$\Sigma^-$& $-{1\over6}-{5\over6}({2\over3}D^2+F^2)$&$-{1\over12}-{5\over12}(D-F)^2$ \\[2mm]
$\Lambda$& $0$&$-{5\over6}DF$ \\[2mm]
$\Sigma^0\Lambda$&${5\over3\sqrt{3}}DF$ &${5\over6\sqrt{3}}DF$ \\[2mm]
\hline
\end{tabular}
\end{center}

\begin{center}
\begin{tabular}{|l|c|c|}\hline
&\multicolumn{2}{|c|}{$\beta^{\prime(X)}_B$}\\[1mm]  \hline
&$\pi$&$K$\\[1mm] \hline
$p$&$-{5\over27}\c^2$ &${5\over108}\c^2$ \\[2mm]
$n$&${5\over27}\c$ &${5\over54}\c^2$ \\[2mm]
$\Xi^-$&$-{5\over54}\c^2$ &$-{5\over108}\c^2$ \\[2mm]
$\Xi^0$ & ${5\over54}\c^2$& ${5\over27}\c^2$\\[2mm]
$\Sigma^+$& ${5\over108}\c^2$&$-{5\over27}\c^2$ \\[2mm]
$\Sigma^-$& $-{5\over108}\c^2$&$-{5\over54}\c^2$ \\[2mm]
$\Lambda$& $0$&${5\over36}\c^2$ \\[2mm]
$\Sigma^0\Lambda$&$-{5\over3\sqrt{3}}\c^2$ &$-{5\over36\sqrt{3}}\c^2$ \\[2mm]
\hline
\end{tabular}
\end{center}

\end{document}